\documentclass[12pt,letterpaper]{article}          %% LaTeX 2e (preferred)
\usepackage{osajnl}
\usepackage[draft]{hyperref} % optional

\begin{document}

\title{Examples of Gaussian cluster computation}

\author{Peter van Loock}
%% for REVTeX4, each author name can be set in a separate \author{} field

\address{National Institute of Informatics (NII),\\ 2-1-2
Hitotsubashi, Chiyoda-ku, Tokyo 101-8430, Japan}

\email{vanloock@nii.ac.jp}

\begin{abstract}
We give simple examples that illustrate the principles of one-way
quantum computation using Gaussian continuous-variable cluster
states. In these examples, we only consider single-mode
evolutions, realizable via linear clusters. In particular, we
focus on Gaussian single-mode transformations performed through
the cluster state. Our examples highlight the differences between
cluster-based schemes and protocols in which special quantum
states are prepared off-line and then used as a resource for the
on-line computation.
\end{abstract}

% REPLACE WITH CORRECT OCIS CODES FOR YOUR ARTICLE
                          % NOTE: \ocis{} IS ALIASED TO \pacs{} BUT MUST
                          % FORMAT THE TERMS CORRECTLY FOR EACH JOURNAL

\maketitle %% NULL FUNCTION WITH LATEX 2e; required for REVTeX4

\section{Introduction}

There are currently two equivalent, but conceptually different
models for quantum computation \cite{NielsenChuang}: the more
conventional circuit model and the so-called {\it cluster-state
model}. In the cluster-state model, proposed by Raussendorf and
Briegel \cite{Raussendorf}, the gates for coupling two or more
qubits during a computation are encoded off-line onto a
sufficiently large multi-qubit cluster state. Once the cluster
state has been created, for the actual computation, only one-qubit
measurements and feedforward are needed. In addition to the
multi-qubit entangling gates, by choosing suitable measurement
bases, arbitrary single-qubit gates can be applied to the initial
state when it propagates through the cluster. This model is
sufficient to describe universal quantum computation.

The idea of using {\it continuous variables}
\cite{Braunstein2005a} for quantum computing was investigated by
Lloyd and Braunstein \cite{LloydBraunstein}. They analyzed how to
realize ``quantum gates'', i.e., unitary transformations,
described by Hamiltonians which are arbitrary polynomials of the
annihilation and creation operators of harmonic oscillators, $\hat
a$ and $\hat a^\dagger$, respectively. Equivalently, these
polynomials can be expressed in terms of the continuous position
and momentum variables, $\hat x$ and $\hat p$, respectively, where
$\hat a = \hat x + i \hat p$. In quantum optics, these observables
represent the amplitude and phase quadratures of the optical
modes, while the operator $\hat a$ ($\hat a^\dagger$) annihilates
(creates) a photon in the corresponding mode. Lloyd and
Braunstein's analysis showed that any such Hamiltonian can be
simulated to an arbitrary accuracy by a combination of
Hamiltonians of up to, at least, cubic order. A Hamiltonian of
cubic or higher order corresponds to a nonlinear transformation of
the mode operators $\hat a$ (or the quadratures $\hat x$ and $\hat
p$), whereas any Hamiltonian up to quadratic order leads to only
linear transformations which include beam splitters and squeezers.
In other words, Lloyd and Braunstein demonstrated that in order to
achieve universality, i.e., to implement {\it any} Hamiltonian, at
least one nonlinear transformation is needed. For this purpose, it
is sufficient to use a nonlinear one-mode transformation. Thus,
{\it any} multi-mode transformation can be realized via a
combination of nonlinear one-mode transformations and linear
multi-mode transformations.

In a very recent work \cite{clusterPRL}, the results of Lloyd and
Braunstein on ``continuous-variable quantum computation'' were
combined with the cluster-state approach to quantum computation.
In quantum optical language, continuous-variable cluster states
are multi-mode Gaussian states which can be prepared by coupling
highly squeezed light modes via a quadratic QND-type interaction
\cite{clusterPRL,Zhang}. There are various ways how to realize
this QND coupling, for instance, by using beam splitters and
squeezers \cite{Braunstein2005}, by employing optical cross-Kerr
interactions \cite{MilburnWalls} in the limit of large intensities
\cite{Grangier}, or by considering cluster states of atomic
ensembles coupled via an optical bus mode through off-resonant
light-atom interactions \cite{Julsgaard,Geremia}. The great
advantage of using Gaussian cluster states as a resource for
cluster computation is that they can be generated in an
unconditional, deterministic fashion. This is in sharp contrast to
the discrete-variable linear-optics schemes for cluster
computation where the non-Gaussian single-photon-based cluster
states can be made only probabilistically
\cite{Nielsen2004,Browne}. However, in principle, the
single-photon-based schemes allow for the creation of perfect
photonic cluster states. Due to the finite squeezing of the
initial modes in Gaussian cluster-state preparation, in this case,
the resulting cluster states will always be imperfect.

Many of the features of qubit-based cluster computation
\cite{Raussendorf} also apply to the continuous-variable version
of cluster computation \cite{clusterPRL}. In particular, in order
to perform an arbitrary multi-mode Gaussian transformation, only
continuous-variable (homodyne) measurements are needed and all
these homodyne detections can be done simultaneously, once the
Gaussian cluster state has been prepared. This {\it parallelism}
is analogous to qubit cluster computation when so-called Clifford
gates are realized via the cluster state. In order to employ
Gaussian cluster states for universal quantum computation, at
least one non-Gaussian measurement is required. However, analogous
to the qubit case, the cluster state itself can still be built
solely through Clifford-type (i.e., in the continuous-variable
case, through Gaussian) operations. After adding a non-Gaussian
measurement to the toolbox, the cluster computation can no longer
be done in parallel and feedforward is needed; the correct choice
of subsequent measurement bases will depend on the results of
earlier measurements. This {\it adaptiveness} is again analogous
to the qubit case when computing non-Clifford gates.

\section{Cluster-computation versus off-line schemes}
\label{clustervsoff}

The essence of cluster-state computation can be understood by
considering a sequence of elementary teleportation circuits by
which quantum information is transmitted through the cluster and
potentially manipulated during each elementary step
\cite{Nielsenreview}. Before the original proposal of one-way
quantum computation via cluster states \cite{Raussendorf}, it was
already recognized that teleportation can be used to perform
quantum gate operations \cite{GottesmanChuang}. The basic idea in
these teleportation-based schemes is that the desired gate
operation is applied off-line to an entangled state. Eventually,
during the actual on-line computation, the gate is applied to an
arbitrary input state via quantum teleportation using that
suitably modified entangled resource. This method turned out to be
very useful for possible optical implementations of quantum
computation, since the off-line gate operations can be done
probabilistically without spoiling the on-line computation. For
instance, qubit nonlinear sign shift gates or continuous-variable
cubic phase gates can be applied non-deterministically to
entangled states of sufficiently many photons \cite{KLM} or to
sufficiently squeezed two-mode squeezed states
\cite{BartlettMunro}, respectively. During the teleportation-based
on-line computation, only linear optics and photon counting plus
feedforward \cite{KLM} or Gaussian operations including
feedforward \cite{BartlettMunro} will then be needed in order to
achieve, respectively, a perfect gate operation
near-deterministically or a near-perfect gate deterministically.

A common feature of the above-mentioned teleportation-based
schemes is that the ``difficult'' operations are performed
off-line. In contrast, in cluster-state computation, though also
describable in terms of quantum teleportation circuits, the
off-line state preparation might be reasonably simple; the cluster
states can be generated solely by means of Clifford-type
entangling operations (controlled phase gates). In the case of
single-photon-based qubit schemes, however, these entangling gates
actually {\it are} the ``difficult'' operations, but they can be
achieved for few-qubit few-photon cluster states using nonlinear
optics \cite{Walther}. For continuous variables, Gaussian
operations (including squeezers) are sufficient to build non-ideal
cluster states as a resource for universal quantum computation
\cite{clusterPRL}.

When implementing cluster-state computation, potentially
``difficult" operations can be absorbed into the detection
process; a single non-Clifford projective measurement is necessary
and sufficient in order to do universal quantum computation via
the Clifford-made cluster states. In the continuous-variable case,
such a non-Clifford measurement corresponds to a non-Gaussian
measurement. Photon counting would be an example for such a
non-Gaussian measurement. Whether and to what extent an initial
quantum state is changed during its propagation through the
cluster then depends on the choice of measurement basis in each
elementary step. Thus, even though the cluster state is fixed and
remains ``untouched" until the on-line computation, a sufficiently
broad set of measurements including non-Clifford (non-Gaussian)
measurements still enables one to realize universal gates. This
feature is different from the ``conventional" teleportation-based
off-line schemes. In the latter, in order to apply a particular
gate during a computation, the off-line resource state must be
prepared correspondingly. In other words, different gates require
different off-line resource states. Of course, by combining
several teleportation-based off-line schemes, where each realizes
a different gate through a different entangled resource state, we
may also achieve universality. However, for a particular gate
sequence or algorithm to be computed, a suitable set of off-line
resource states must be selected and used in accordance to the
desired computation. As a consequence, the actual set of off-line
resource states would again be different in every computation, as
opposed to a fixed (sufficiently large) cluster state which can be
used for different cluster-computations.

Hence, cluster-state computation and off-line schemes, though both
based upon quantum teleportation on an elementary level, somewhat
differ in various aspects. One such aspect is universality, i.e.,
whether {\it any} unitary gate can be realized using a given
cluster state or off-line resource state without ever changing
these states. Another aspect might be the different ``degrees of
difficulty" in performing the off-line and on-line operations when
cluster-state or off-line computation are implemented. We will
give examples for this later in the context of continuous
variables. Furthermore, somewhat related with the above criteria,
different types of input states are usually considered for
cluster-state computation and for off-line gate operations.
Whereas arbitrary input states, coming independently from outside,
can be teleported in an off-line scheme, in cluster-state
computation, typically a fixed blank state which is part of the
cluster plays the role of the input.

In the following sections, we will illustrate the similarities and
differences of cluster-based and off-line computation by looking
at very simple continuous-variable examples. In particular, we
only consider the evolution of a single mode, i.e., single-mode
gates. Our main focus will be on Gaussian single-mode
transformations.

\section{Teleportation circuits for continuous-variable cluster-computation}

\begin{figure}[htbp]
\centering
\includegraphics[width=11cm]{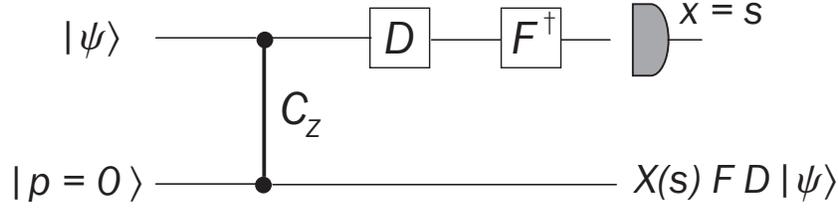}
\caption{Elementary teleportation circuit for continuous-variable
cluster computation.} \label{fig1}
\end{figure}

Qubit cluster-state computation basically relies upon a
combination of one-qubit teleportation circuits
\cite{Nielsenreview,Zhou}. These circuits enable one to teleport
operations diagonal in the computational basis onto an initial
state just by performing measurements on the given cluster state.
The continuous-variable analogue \cite{clusterPRL} of the
one-qubit teleportation circuit is shown in Fig.~\ref{fig1}.

In this elementary teleportation circuit, the ``input state" of
mode 1, $|\psi\rangle$, is coupled to a zero-momentum eigenstate
of mode 2, $|p=0\rangle = \int dx |x\rangle/\sqrt{\pi}$, via a
continuous-variable controlled-$Z$ gate, $C_Z = \exp(2 i \hat x
\otimes \hat x)$. Further, we use the Fourier transform, $F =
\exp[i(\pi/2)(\hat x^2 + \hat p^2)]$, where $F|x\rangle = \int dy
e^{2ixy} |y\rangle/\sqrt{\pi} = |p = x\rangle$, and an arbitrary
operator diagonal in the computational (position) basis, $D =
\exp[i f(\hat x)]$. The measurement of mode 1 is supposed to be a
measurement of the position observable $\hat x$ with a classical
outcome $s$. This classical result then appears in the
``teleported state" of mode 2 through the position displacement
operator $X(s) = \exp(-2i s \hat p)$, where $X(s)|x\rangle =
|x+s\rangle$.

The most important feature of the elementary teleportation circuit
is that the $C_Z$ gate commutes with any diagonal operator $D$.
Therefore, we can consider an equivalent scheme where the operator
$D$ acts upon the input state $|\psi\rangle$ {\it before} the
$C_Z$ gate is applied to the two modes. In other words, the
circuit in Fig.~\ref{fig1} is identical to a circuit where the
state $|\psi'\rangle = D |\psi\rangle$ is teleported onto mode 2
and the only operation between the $C_Z$ gate and the $x$-homodyne
detector of mode 1 is the inverse Fourier transform $F^\dag$.
Using $|\psi'\rangle = \int dx \psi'(x) |x\rangle$ as the ``input
state" of mode 1, directly after the $C_Z$ gate, we obtain the
two-mode state,
\begin{eqnarray}
C_Z\, |\psi'\rangle \otimes |p=0\rangle =
\frac{1}{\sqrt{\pi}}\,\int dx dy \,\psi'(x)\, e^{2i x y}\,
|x\rangle |y\rangle\,,
\end{eqnarray}
where we also used $|x\rangle |y\rangle\equiv |x\rangle\otimes
|y\rangle$. Now absorbing the inverse Fourier transform into the
$x$-homodyne detector means that we will apply the projector
$F|x\rangle\langle x|F^\dag = |p=x\rangle\langle p=x|$ onto mode 1
and effectively project it onto the $p$-basis $\{F|x\rangle\} =
\{|p=x\rangle\}$. For a measurement result $p=s$, the conditional
state of mode 2 becomes
\begin{eqnarray}
\frac{1}{\sqrt{\pi}}\,\int dx dy \,\psi'(x)\, e^{2i x (y - s)}\,
|y\rangle = X(s) F |\psi'\rangle\,,
\end{eqnarray}
in agreement with Fig.~\ref{fig1}. The crucial point in this
circuit is that, now again equivalently assuming that the operator
$D$ acts between the $C_Z$ gate and the $p$-homodyne detection,
this operation $D$ can be also absorbed into the detector. Thus,
for a given state $|\psi\rangle$ of mode 1, the ``output state" of
mode 2, $X(s) F D |\psi\rangle$, can be manipulated solely by
choosing different measurement bases $\{D^\dag |p\rangle\}$
corresponding to the observables $D^\dag\hat p D$. By
concatenating these elementary teleportation circuits one can
generate output states of the form $\cdots X(s_2) F D_2 X(s_1) F
D_1 |\psi\rangle$ and hence realize any single-mode unitary gate
\cite{clusterPRL,LloydBraunstein}. Thereby it depends on the type
of the desired operations $D_j$ (Clifford or non-Clifford) how
easily the position-displacements $X(s_j)$ can be commuted through
in order to be corrected. Moreover, the momentum-squeezed resource
states will always be finitely squeezed (instead of being
infinitely squeezed, unphysical zero-momentum eigenstates). This
ultimately leads to distortions of the states that are teleported
through the cluster \cite{clusterPRL}. These distortions might be
suppressed for the case of Clifford (Gaussian) operations by
exploiting parallelism and postselection \cite{clusterPRL}.

Let us finally mention that an elementary teleportation circuit
analogous to the one in Fig.~\ref{fig1} can be constructed where
the resource state is a zero-position eigenstate $|x=0\rangle$,
the operator $D$ is diagonal in the $p$ basis, $D = \exp[i f(\hat
p)]$, the controlled phase gate has the form $\exp(2i\hat
p\otimes\hat p)$, the measurement after the inverse Fourier
transform is a $p$ measurement with classical result $p=t$, and
the output state contains a momentum-displacement $Z(-t) =
\exp(-2i t \hat x)$, where $Z(t)|p\rangle = |p+t\rangle$.

\section{Linear clusters for single-mode evolutions}

\begin{figure}[t]
\centering
\includegraphics[width=13cm]{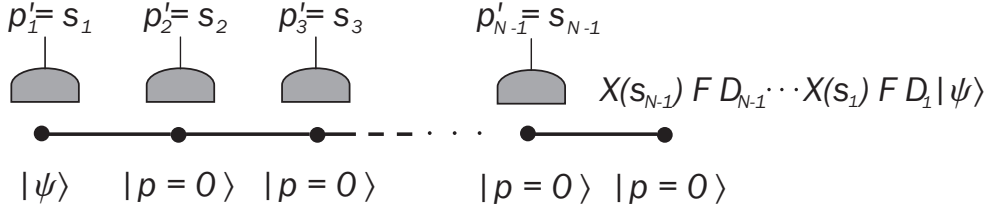}
\caption{Using linear Gaussian cluster states for single-mode
evolutions: a chain of coupled squeezed modes is generated via
controlled-$Z$ gates; an ``input state" $|\psi\rangle$ can be
teleported through the cluster and potentially manipulated at each
step; the form of the output state depends on the choice of
measurement bases for detecting the observables $\hat
p'_j=D_j^\dag\hat p_j D_j$.} \label{fig2}
\end{figure}

The teleportation circuit described in the preceding section is
the elementary building block for any cluster computation
including both single-mode and multi-mode gates
\cite{Nielsenreview,clusterPRL}. In general, in order to apply a
multi-mode gate via a cluster state, the input state must
propagate through a nonlinear cluster. In the continuous-variable
case, highly squeezed light modes can be coupled via QND-type
controlled-$Z$ operations in order to create the required graph
state \cite{clusterPRL}. In the following, however, we will focus
on linear cluster states where each node has at most two links.
Linear cluster states are a sufficient resource to implement
arbitrary single-qubit or single-mode transformations. In the
continuous-variable case, for some input state of mode 1,
$|\psi\rangle$, an arbitrary single-mode evolution can be realized
by teleporting the state through a linear chain of modes which are
all coupled via $C_Z$ gates just as the two modes in the
elementary two-mode teleportation circuit (see Fig.~\ref{fig2}).

When using the linear $N$-mode cluster state in Fig.~\ref{fig2}
for single-mode transformations, at each step $j$, an appropriate
observable $\hat p'_j = D_j^\dag\hat p_j D_j$ must be measured
yielding a set of $N-1$ classical results $p'_j = s_j$. The
corresponding position-displacements $X(s_j)$ which appear in the
output state of mode $N$ can be corrected at the end. For Clifford
operations, by definition, we can always write
\begin{equation}
D_j U_{\rm WH} = U'_{\rm WH} D_j\,,
\end{equation}
where $D_j$ is the desired Clifford group operation, $U_{\rm WH}$
is a Weyl-Heisenberg (WH) group transformation [a phase-space
displacement such as $X(s)$ or $Z(t)$] effected through previous
cluster computations, and $U'_{\rm WH}$ is a modified WH group
transformation which can be undone for correcting the output
state. Note that the desired operation $D_j$ remains unchanged
here and only the WH transformation is modified. In this case, for
example, we obtain $\cdots X(s_2) F D_2 X(s_1) F D_1 |\psi\rangle
= \cdots X(s_2) U'_{\rm WH}(s_1) F D_2 F D_1 |\psi\rangle$, if
$D_2$ is an element of the Clifford group (the Fourier transform
also belongs to the Clifford group). The WH transformations can
then be corrected or further commuted through.

However, in the case of non-Clifford (non-Gaussian) operations
$D_j$, the correct choice of measurement bases depends on the
outcome of previous measurements, because only via this
feedforward is it possible to commute through and correct the
position displacements. As a result, the parallelism for Clifford
operations, i.e., the feature that all measurements can be
performed simultaneously, no longer holds for non-Clifford gates.

In the non-Clifford case, we have
\begin{equation}
D'_j U_{\rm WH} = U_{\rm WH} D_j\,,
\end{equation}
where now the desired non-Clifford operation $D_j$ can be, in
general, only applied to the output state if the measurement basis
is chosen such that in the current cluster computation, a modified
$D'_j$ will be added to the incoming states. In general, this
modified $D'_j$ will depend on the measurement results of the
previous cluster computations (contained in $U_{\rm WH}$). More
specifically, when combining two elementary cluster circuits
(using a linear 3-mode cluster state), the output state becomes
\begin{eqnarray}
|\psi\rangle_{\rm out} &=& X(s_2) F D_2' X(s_1) F D_1|\psi\rangle
\,,
\end{eqnarray}
where $D_1$ is due to the first circuit with a measurement of
$D_1^{\dagger}\hat p D_1$ and $D_2'\equiv D_2'(\kappa,s_1)$ is due
to the second circuit with a measurement of $D_2'^{\dagger}\hat p
D_2'$. This can be rewritten as
\begin{eqnarray}
|\psi\rangle_{\rm out} &=& X(s_2) Z(s_1) F D_2 F D_1|\psi\rangle
\,,
\end{eqnarray}
with $D_2'(\kappa,s_1) X(s_1) = X(s_1) D_2(\kappa)$ for a
(desired) non-Clifford group operation $D_2\equiv D_2(\kappa)$.
For example, in order to effect the cubic gate
$D_2(\kappa)=e^{i\kappa\hat x^3}$, we have to measure the
observable $D_2'^{\dagger}(\kappa,s_1)\hat p D_2'(\kappa,s_1)$ of
mode 2 with $D_2'(\kappa,s_1)=e^{3i\kappa s_1 \hat x(s_1 - \hat
x)}D_2(\kappa)$.

\section{Gaussian single-mode evolutions: single-mode squeezer}

In this section, we will now discuss an example for implementing a
single-mode evolution using a linear Gaussian cluster state. Apart
from the Gaussian cluster state, we also assume that all the
measurements will be Gaussian measurements corresponding to
homodyne detections. In other words, the argument of the diagonal
operators $D$ will be, at most, of quadratic order in $\hat x$.
Since an operator $D$ that contains a linear function of $\hat x$
is a momentum-displacement operator, measuring $D^\dag\hat p D$ in
this case simply means that the measured $p$ value must be shifted
correspondingly. In our example, we will use operators of
quadratic form, $D=e^{i\kappa\hat x^2}$. With these operators, we
can realize a single-mode squeezer. A single-mode squeezer is an
important primitive for performing Gaussian transformations. In
fact, {\it any} multi-mode Gaussian transformation can be
decomposed into a passive linear-optics network, a set of
single-mode squeezers, and another linear-optics circuit
\cite{Braunstein2005}. When implementing a single-mode squeezing
transformation via a Gaussian cluster state, we encounter the
typical feature of Clifford parallelism. Moreover, via a
sufficiently long but fixed cluster state we may still achieve
universal squeezing.

In our example of a single-mode squeezer, the desired total
Clifford group operation acting on the state $|\psi\rangle$ is
described by $U = e^{r(\hat a^2 - \hat a^{\dagger 2})/2}= e^{i r
(\hat x \hat p + \hat p \hat x)}$ with a squeezing parameter $r$.
We may realize this operation by combining four elementary cluster
circuits using a linear five-mode cluster state (see
Fig.~\ref{fig2} with $N=5$). In each step $j=1...4$, the operation
$X(s_j) F D_j$ with $D_j\equiv D(\kappa_j)\equiv e^{i\kappa_j \hat
x^2}$ is applied. As a result, after correcting the phase-space
displacements $X(s_j)$, the output state of mode 5 will be $F D_4
F D_3 F D_2 F D_1 |\psi\rangle$. Now by choosing $\kappa_1 =
\kappa_2 = \kappa$, we obtain $F D_2 F D_1 = F^2 e^{i\kappa \hat
p^2} e^{i\kappa \hat x^2}$. Up to rotations and higher-order terms
in $\kappa$, this already corresponds to a single-mode squeezing
operation, since $e^{i\kappa \hat p^2} e^{i\kappa \hat x^2} =
e^{i\kappa (\hat x^2 + \hat p^2)}e^{i \kappa^2 (\hat x \hat p +
\hat p \hat x)/2} + {\rm O}(\kappa^3)$. However, by further
applying $F D_4 F D_3 = F^2 e^{-i\kappa \hat p^2} e^{-i\kappa \hat
x^2}$ with $\kappa_3 = \kappa_4 = -\kappa$, we can add more
squeezing and at the same time undo the unwanted rotations,
because now we have $e^{-i\kappa \hat p^2} e^{-i\kappa \hat x^2} =
e^{i \kappa^2 (\hat x \hat p + \hat p \hat x)/2}e^{-i\kappa (\hat
x^2 + \hat p^2)} + {\rm O}(\kappa^3)$. Thus, the effective total
operation will be
\begin{eqnarray}
F D_4 F D_3 F D_2 F D_1 = F^2 e^{-i\kappa \hat p^2} e^{-i\kappa
\hat x^2} F^2 e^{i\kappa \hat p^2} e^{i\kappa \hat x^2} = e^{i
\kappa^2 (\hat x \hat p + \hat p \hat x)} + {\rm O}(\kappa^3)\,,
\end{eqnarray}
up to a global sign. This is approximately a single-mode squeezer
with squeezing $r\equiv \kappa^2$. It leads to some squeezing in
the position, $\hat x_{\rm out}=(1-\kappa^2)\hat x + {\rm
O}(\kappa^3)$, and the corresponding antisqueezing in the
momentum, $\hat p_{\rm out}=(1+\kappa^2)\hat p + {\rm
O}(\kappa^3)$.

In the above protocol, the measurement bases do not change
depending on the results of earlier measurements, because the
desired output operation is a Clifford group operation. We
explicitly demonstrate this for modes 1 and 2.

When combining two elementary clusters, as described before, and
detecting $D_j^{\dagger}\hat p_j D_j$ of modes 1 and 2 with
measurement results $s_1$ and $s_2$, mode 3 will be projected onto
the state $X(s_2) F D_2 X(s_1) F D_1|\psi\rangle$. Since $D_2$ is
an element of the Clifford group, the correction based on the
result of the first measurement, $X(s_1)$, can be commuted through
the $D_2$-operation, $D_2 X(s_1) = Z(\kappa s_1)X(s_1)D_2$ (up to
a global phase). Note that here the desired Clifford group
operation $D_2$ remains unchanged, while the WH group elements
necessary for correction become modified. This is why also the
detection of mode 2 is just a measurement of $D_2^{\dagger}\hat
p_2 D_2$ independent of the result of the first measurement of
mode 1. So finally, the resulting state
\begin{eqnarray}
|\psi'\rangle &=& X(s_2) F Z(\kappa s_1) X(s_1)
D_2 F D_1|\psi\rangle \nonumber\\
&=& X(s_2-\kappa s_1) Z(s_1) F^2 e^{i\kappa \hat p^2} e^{i\kappa
\hat x^2}|\psi\rangle
\end{eqnarray}
can be corrected to obtain the desired transformation, as
described above. The remaining operations in order to complete the
single-mode squeezer can be similarly corrected.

In the scheme described above, measuring the observables $\hat
p'_j=D_j^\dag\hat p_j D_j$ (instead of $\hat p_j$) for $D_j\equiv
e^{i\kappa_j \hat x^2}$ means that a linear combination of
position and momentum should be detected, $\hat p'_j = \hat p_j +
\kappa_j\hat x_j$. This, however, simply corresponds to the
measurement of rotated quadratures, $(\hat p_j\cos\theta_j - \hat
x_j\sin\theta_j)/(\cos\theta_j)$ with $\kappa_j\equiv
-\tan\theta_j$, where the measurement results must be rescaled by
the factor $\cos\theta_j$. Thus, solely by adjusting the local
oscillator phase $\theta_j=\tan^{-1}(-\kappa_j)$ of the homodyne
detectors, we can measure any observable $\hat p'_j=D_j^\dag\hat
p_j D_j$ and no additional squeezers are required in front of the
detectors. This property, of course, does not only apply to the
cluster scheme for the single-mode squeezer. {\it Any} multi-mode
Gaussian transformation can be performed via a given cluster state
solely by doing suitable homodyne measurements \cite{clusterPRL}.

In the realistic case, finitely squeezed resources are used in
order to build the linear five-mode cluster state. When realizing
the single-mode squeezing transformation via this finitely
squeezed cluster state, at each of the four measurement steps
extra noise will be added to the output state, thus distorting the
desired squeezing transformation appearing in mode 5. However,
since the desired operation is a Clifford group operation,
parallelism and postselection could be used to suppress the effect
of these distortions \cite{clusterPRL}. In this case, modes 3 and
4 are measured {\it before} the $2-5$ cluster state is attached to
mode 1 which is in the input state $|\psi\rangle$. For the final
attachment and detections of modes 1 and 2, only those $2-5$
cluster states will be postselected which lead to the smallest
distortions of the transformed input state. The efficiency of this
method depends on the form of the input state $|\psi\rangle$ and
hence on the encoding used in the cluster computation
\cite{clusterPRL}.

Using a linear Gaussian cluster state, we may now apply some
squeezing $r\equiv \kappa^2$ to the input state $|\psi\rangle$.
For a sufficiently long linear cluster state, we may repeat this
procedure and gradually add more squeezing. However, the degree of
output squeezing in $|\psi\rangle_{\rm out}$ can be controlled
solely through the choice of the detection bases without ever
changing the cluster state. For example, if we decide to apply
only little squeezing to the state $|\psi\rangle$, most of our
measurements will be simple $p$-detections in order to propagate
the state through the cluster. Different squeezing transformations
can be performed using the same fixed cluster state but different
homodyne measurements. This kind of universality is the typical
feature of cluster-computation. In the next section, we will
compare this and other properties of cluster-computation with
those of ``off-line schemes".

\section{Comparison to off-line schemes}

\begin{figure}[t]
\centering
\includegraphics[width=10cm]{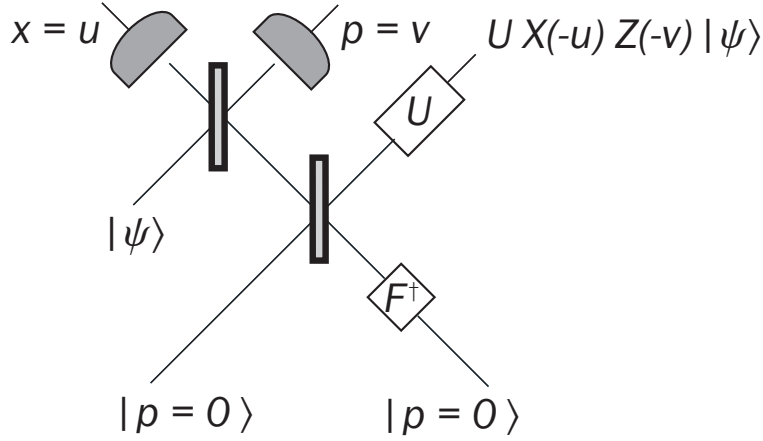}
\caption{Off-line transformation of a quantum state $|\psi\rangle$
via continuous-variable quantum teleportation using a suitably
modified two-mode squeezed state (here in the limit of infinite
squeezing).} \label{fig3}
\end{figure}

Various continuous-variable schemes have been proposed in which
particular transformations (gates) are applied to some input state
via a suitably prepared off-line resource state
\cite{BartlettMunro,GKP,Filip,Lance}. In a teleportation-based
continuous-variable off-line scheme, similar to that for discrete
variables and single-photon states \cite{KLM}, an entangled
two-mode squeezed state is modified beforehand in order to apply a
desired transformation to some input state via quantum
teleportation \cite{BartlettMunro}. In the limit of infinite
squeezing, the two-mode squeezed state corresponds to the output
of a symmetric beam splitter with the two inputs $|p=0\rangle$ and
$F^\dag |p=0\rangle = |x=0\rangle$ (see Fig.~\ref{fig3}). Note
that the infinitely squeezed two-mode squeezed state $\int dx
|x\rangle |x\rangle /\sqrt{\pi}$ is equivalent to a two-mode
cluster state, $e^{2i\hat x \otimes \hat x}|p=0\rangle |p=0\rangle
= \int dx |p=x\rangle |x\rangle /\sqrt{\pi}$ up to a local Fourier
transform.

Quantum teleportation of an input state $|\psi\rangle$ is achieved
by combining the input mode with one half of the two-mode squeezed
state (mode 1) at a second symmetric beam splitter
\cite{cvteleportation} (Fig.~\ref{fig3}). Via $x$ and $p$ homodyne
detections at the two output ports, the linear combinations
$u=x_{\rm in} - x_1$ and $v=p_{\rm in} + p_1$ can be determined.
Depending on these classical measurement results, the conditional
state of mode 2 (the other half of the entangled state) becomes a
replica of the input state $|\psi\rangle$ up to known position and
momentum displacements, $X(-u) Z(-v) |\psi\rangle$
\cite{footnote}.

Now consider the situation where the position and momentum
displacements in the output state $X(-u) Z(-v) |\psi\rangle$ are
not corrected and another unitary transformation $U$ is applied to
the state of mode 2, $U X(-u) Z(-v) |\psi\rangle$. This scenario
is equivalent to a scheme in which the unitary transformation $U$
is performed {\it before} any measurements are done. In other
words, the two-mode squeezed state gets modified before it is used
as a resource for quantum teleportation, $\int dx |x\rangle
|x\rangle /\sqrt{\pi} \rightarrow (\mbox{1$\!\!${\large 1}}\otimes
U)\int dx |x\rangle |x\rangle /\sqrt{\pi}$. Now if $U$ is an
element of the Clifford group, we can again write $U X(-u) Z(-v) =
U'_{\rm WH}(u,v) U$ where $U'_{\rm WH}$ is a modified WH group
transformation, but the desired $U$ is unchanged. Thus, after
correcting the known displacements $U'_{\rm WH}$, we can generate
the output state $U |\psi\rangle$. In this way, quantum gates can
be applied on-line via quantum teleportation
\cite{GottesmanChuang} after the corresponding gate has been
applied {\it off-line} to the entanglement resource. The scheme
described here is just the continuous-variable version of this
method \cite{BartlettMunro}. For example, if the desired
transformation $U$ is a single-mode squeezer that generates $\hat
x_{\rm out} = e^{-r}\hat x$ and $\hat p_{\rm out} = e^{+r}\hat p$,
then we have $U X(-u) Z(-v) |\psi\rangle = X(-e^{-r}u) Z(-e^{+r}v)
U |\psi\rangle$ which can be easily corrected to obtain $U
|\psi\rangle$.

Note that in order to preserve the set of tools needed in the
actual on-line teleportation scheme (homodyne detections and
displacements), it is necessary that $U$ is an element of the
Clifford group. If $U$ is not an element of the Clifford group, we
would have to apply a different $U'$ to the entanglement resource
in order to be able to correct the teleported state via simple
phase-space displacements and still effect the desired gate $U$,
since then $U' U_{\rm WH} = U_{\rm WH} U$. However, in this case,
$U'$ will depend on the classical results of the homodyne
detections $u$ and $v$ during the on-line teleportation protocol
(contained in $U_{\rm WH}$). As a result, we can no longer apply
the correct $U'$ off-line to the entanglement resource {\it
before} the on-line teleportation. The complications that arise
here in the non-Clifford case are similar to those in
cluster-state computation. However, in cluster computation, we can
still apply the correct non-Clifford $D_j$ in every step $j$,
because this can be achieved by choosing the right measurement
basis depending on the measurement results in {\it previous}
steps.

In order to obey the rules of ``off-line computation", we must
apply the desired gate operation $U$ off-line before the on-line
teleportation. Therefore, the resulting teleported state will
always be of the form $U X(-u) Z(-v) |\psi\rangle$. Now even if
$U$ is not an element of the Clifford group, it can be commuted
through the phase-space displacements. However, in this case, the
correction operation will become more complicated than simple
displacements. In general, if $U_k$ describes an interaction of
$k$th order in $\hat x$ and $\hat p$, we have $U_k U_{\rm WH} =
U_{k-1} U_k$ ($k>1$) where $U_{k-1}$ is a known interaction of
$(k-1)$th order depending on the form of $U_{\rm WH}$
\cite{BartlettMunro}. Thus, if we allow for more complicated
correction operations than simple phase-space displacements, we
may still teleport a non-Clifford operation onto the state
$|\psi\rangle$. For example, a cubic gate $U_3$ would require a
correction operation of the form $U_2$ which is a Clifford
(Gaussian) operation and includes displacements and squeezers
\cite{BartlettMunro}.

In Sec.~\ref{clustervsoff}, we mentioned various aspects in which
cluster computation and off-line schemes differ. For example, the
essence of implementing a teleportation-based off-line scheme is
that the ``difficult'' operations are performed off-line. In the
above continuous-variable protocol such a ``difficult'' operation
$U$ is applied off-line to a two-mode squeezed state. This
off-line operation could be, for instance, a cubic phase gate,
while during the on-line teleportation protocol Gaussian
operations are sufficient. Compare this with a cubic gate realized
via a continuous-variable cluster state. In this case, Gaussian
(Clifford) operations suffice in order to build the cluster
off-line. However, a non-Gaussian measurement is required on-line
to accomplish the cubic gate. This could be realized, for example,
by putting a cubic gate in front of the $p$-homodyne detector.

When implementing Clifford gates for continuous variables,
off-line and cluster-based schemes compare as follows. Consider as
the desired gate a single-mode squeezing transformation. In the
off-line scheme, assume that we are given a highly squeezed
two-mode squeezed state for free. Now the ``difficult'' operation
would be to apply the single-mode squeezing transformation $U$
off-line to the two-mode squeezed state. Eventually, this
single-mode squeezing $U$ can be teleported onto some input state
via fixed homodyne detections and conditional phase-space
displacements. In contrast, given a highly squeezed linear
Gaussian cluster state, no further manipulation of this resource
is needed off-line. However, for the on-line measurements, as
described in the preceding section, we have to detect ``squeezed
quadratures'', i.e., observables of the form $D^\dag\hat p D$
where $D$ represents a quadratic gate. This could be realized by
putting a single-mode squeezer (plus phase shifters) in front of
the $p$-homodyne detector. Of course, such an approach would be
awkward, because the measurement of $D^\dag\hat p D$ in this case
only requires adjustment of the homodyne detector's local
oscillator phase. However, this example illustrates the conceptual
difference between the off-line and the cluster scheme: in the
former, squeezing is applied off-line depending on the desired
output squeezing and the homodyne measurements are fixed; in the
latter, the cluster state is fixed and the measured observables
are ``squeezed'' depending on the desired output squeezing.

As a consequence, in cluster computation, in general, deciding the
final transformation can be postponed until the very end. This
leads to another conceptual difference between cluster and
off-line schemes, namely universality. What kind of gates can be
realized via a given cluster or off-line state? If the cluster
state is sufficiently large, {\it any} gate or gate sequence is
realizable via the same cluster state only by adjusting the
measurement bases. An off-line entangled state must be
correspondingly modified for implementing particular gates and
different off-line states must be selected for different
computations. For example, when performing the single-mode
squeezing transformation via a (sufficiently long) linear Gaussian
cluster state, we may still decide how much squeezing we apply
{\it after} the cluster state has been created. In the case of a
teleportation-based off-line scheme, we have to apply the
corresponding amount of squeezing to the entangled state (or
select a suitable set of modified entangled states) {\it before}
the on-line protocol.

Finally, let us compare the different types of input states in
cluster and off-line computation. Typically, in cluster
computation, the ``input state'' will be some fixed blank state
which is part of the cluster. In contrast, in the
teleportation-based off-line protocol, an arbitrary input state is
coming independently from outside and is coupled to one half of
the entangled resource state only during the on-line computation.
This coupling will be performed only provided the resource state
has been suitably prepared. Of course, one may consider a
situation where also a cluster state is only attached to some
input state (which could be part of some other cluster state)
during the on-line computation. In fact, such a scenario turns out
to be useful when implementing Gaussian operations via Gaussian
cluster states due to the Clifford parallelism \cite{clusterPRL}.
In this case, some of the homodyne measurements can be made {\it
before} the cluster state is attached to the input state or
another cluster state. Similar to off-line computation schemes,
only those cluster states would be postselected which are best
suited to effect the desired transformation (and least likely to
distort the input state).

\section{Conclusion}

In summary, we briefly described the concept of
continuous-variable cluster computation using Gaussian cluster
states. As an example of Gaussian cluster computation, we
considered a single-mode squeezer. This example is particularly
simple, because it represents a single-mode evolution which can be
realized via linear cluster states. Moreover, the desired
operation in this case is a Clifford (Gaussian) operation. As a
result, during the computation, the measurement bases do not
change depending on the results of earlier measurements. In this
case, all measurements in the cluster computation can be performed
at the same time. Finally, we discussed some conceptual
similarities and differences between cluster-computation and
off-line computation and illustrated these for the case of
continuous variables.

\section*{Acknowledgments}

I would like to thank Mile Gu, Nicolas C. Menicucci, Michael A.
Nielsen, Timothy C. Ralph, and Christian Weedbrook for their
collaboration with me on continuous-variable cluster computation
and Kae Nemoto for useful discussions. I acknowledge funding from
MIC in Japan.


\begin{thebibliography}{99}
\bibitem{footnote} usually, the quantum optical phase-space
displacement operator $D(\alpha) = \exp(\alpha \hat a^\dag -
\alpha^* \hat a)$ is used to describe the displacements in
``Alice's" measurement basis and ``Bob's" correction operation for
continuous-variable quantum teleportation. Here we prefer to stick
to the WH group elements, where $Z(v)X(u)$ differs from
$D(\alpha)$ only by a global phase with $\alpha = u + i v$.
Alice's measurement basis is then described by
$[Z(v)X(u)\otimes\mbox{1$\!\!${\large 1}}]\int dx |x\rangle
|x\rangle /\sqrt{\pi}$. Bob's correction operation would be
$Z(v)X(u)$ in this case. In a communication rather than
computation scenario where Alice and Bob are spatially separated,
of course, Alice needs to send the results $u$ and $v$ to Bob via
a classical channel.
\end{thebibliography}
\end{document}